\documentclass[10pt,conference]{IEEEtran}

\usepackage{epsfig}
\usepackage{graphicx}
\usepackage{amsmath}
\usepackage{amssymb}
\usepackage{float}
\usepackage{url}
\newtheorem{definition}{Definition}

\newtheorem{proposition}{Proposition}
\newtheorem{corollary}{Corollary}
\newtheorem{theorem}{Theorem}

\newtheorem{remark}{Remark}

\newcommand{\reals}{{\rm I\!R}}
\newcommand{\expectation}{{\rm I\!E}}

\newenvironment{proof}[1][Proof]{\begin{trivlist}
\item[\hskip \labelsep {\bfseries #1}]}{\end{trivlist}}

\begin{document}

\title{Tight Bounds for Symmetric Divergence Measures and a New Inequality
Relating $f$-Divergences}

\author{\IEEEauthorblockN{Igal Sason\\
\hspace*{-0.4cm} Department of Electrical Engineering\\
Technion, Haifa 32000, Israel\\
E-mail: \url{sason@ee.technion.ac.il}\\}}

\maketitle

\begin{abstract}
Tight bounds for several symmetric divergence measures are introduced, given in terms of the total variation distance.
Each of these bounds is attained by a pair of 2 or 3-element probability distributions.
An application of these bounds for lossless source coding is provided, refining and improving a certain
bound by Csisz\'{a}r. A new inequality relating $f$-divergences is derived, and its use is exemplified.
The last section of this conference paper is not included in the recent journal paper \cite{Sason_IT15},
as well as some new remarks that are linked to new references.
\end{abstract}

\section{Introduction and Preliminaries}
\label{section: Introduction}
Divergence measures are widely used in information theory, machine learning, statistics,
and other theoretical and applied branches of mathematics (see, e.g., \cite{Basseville},
\cite{CsiszarS_FnT}, \cite{ReidW11}).
The class of $f$-divergences forms an important class of divergence measures. Their properties,
including relations to statistical tests and estimators, were studied, e.g., in
\cite{CsiszarS_FnT} and \cite{LieseV_IT2006}.

In \cite{Gilardoni06}, Gilardoni studied the problem of minimizing an arbitrary
{\em symmetric} $f$-divergence for a given total variation distance (these terms
are defined later in this section), providing a closed-form solution of this
optimization problem. In a follow-up paper by the same author \cite{Gilardoni10},
Pinsker's and Vajda's type inequalities were studied for symmetric $f$-divergences,
and the issue of obtaining lower bounds on $f$-divergences for a fixed total
variation distance was further studied.

One of the main results in \cite{Gilardoni10} was a further derivation of a simple closed-form
lower bound on the relative entropy in terms of the total variation distance. The
relative entropy is an asymmetric $f$-divergence, as it is clarified in the continuation
to this section. The lower bound on the relative entropy suggests an improvement over
Pinsker's and Vajda's inequalities. A derivation of a simple and reasonably
tight closed-form upper bound on the infimum of the relative entropy has been
also provided in \cite{Gilardoni10} in terms of the total variation distance.
An exact characterization of the minimum of the relative entropy subject to a fixed
total variation distance has been derived in \cite{FedotovHT_IT03} and \cite{Gilardoni06}.

Sharp inequalities for $f$-divergences were recently studied in \cite{GSS_IT14}
as a general problem of maximizing or minimizing an arbitrary $f$-divergence between two probability
measures subject to a finite number of inequality constraints on other $f$-divergences.
The main result stated in \cite{GSS_IT14} is that such infinite-dimensional
optimization problems are equivalent to optimization problems over finite-dimensional
spaces where the latter are numerically solvable.

The total variation distance has been further studied from an information-theoretic perspective
by Verd\'{u} \cite{Verdu-ITA14}, providing upper and lower bounds on the total variation
distance between two probability measures $P$ and $Q$ in terms of the distribution of
the relative information $\log \frac{dP}{dQ}(X)$ and $\log \frac{dP}{dQ}(Y)$ where
$X$ and $Y$ are distributed according to $P$ and $Q$, respectively.

Following previous work, {\em tight} bounds on symmetric $f$-divergences and related distances
are introduced in this paper.
An application of these bounds for lossless source coding is provided,
refining and improving a certain bound by Csisz\'{a}r \cite{Csiszar67b}.
The material in this conference paper appears in the recently published journal
paper by the same author \cite{Sason_IT15}. However, we also provide in this conference
paper a new inequality relating $f$-divergences, and its use is exemplified; this material
is not included in the journal paper \cite{Sason_IT15} since it does not necessarily refer
to symmetric $f$-divergences.

The paper is organized as follows: tight bounds for several symmetric divergence
measures, which are either symmetric $f$-divergences or related symmetric distances,
are introduced without proofs in Section~\ref{section: Derivation of Tight Bounds on Symmetric Distance Measures};
these bounds are expressed in terms of the total variation distance.
An application for the derivation of an improved and refined bound in
the context of lossless source coding is provided in Section~\ref{section: bound for lossless source coding}.
The full version of this work, including proofs of the tight bounds in Section~\ref{section: bound for lossless source coding},
appears in \cite{Sason_IT15}.
Section~\ref{section: a new inequality relating f-divergences} provides a new inequality that relates between
$f$-divergences; this inequality is proved since it is not included in the journal paper \cite{Sason_IT15}.

We end this section by introducing some preliminaries.
\begin{definition}
Let $P$ and $Q$ be two probability distributions with a
common $\sigma$-algebra $\mathcal{F}$.
The {\em total variation distance} between $P$ and $Q$ is
$d_{\text{TV}}(P, Q)  \triangleq \sup_{A \in \mathcal{F}} |P(A) - Q(A)|.$
\label{definition: total variation distance}
\end{definition}
If $P$ and $Q$ are defined on a countable set, it is simplified to
\begin{equation}
d_{\text{TV}}(P, Q) = \frac{1}{2} \sum_{x} \bigl|P(x) - Q(x)\bigr| =
\frac{||P-Q||_1}{2}.
\label{eq: the L1 distance is twice the total variation distance}
\end{equation}

\begin{definition}
Let $f \colon (0, \infty) \rightarrow \reals$ be a convex function with $f(1)=0$,
and let $P$ and $Q$ be two probability distributions.
The {\em $f$-divergence} from $P$ to $Q$ is defined by
\begin{equation}
D_f(P||Q) \triangleq \sum_{x} Q(x) \, f\left(\frac{P(x)}{Q(x)}\right)
\label{eq:f-divergence}
\end{equation}
with the convention that
\begin{align*}
&\ 0 f\Bigl(\frac{0}{0}\Bigr) = 0, \quad
f(0) = \lim_{t \rightarrow 0^+} f(t), \\
&\ 0 f\Bigl(\frac{a}{0}\Bigr) = \lim_{t \rightarrow 0^+}
t f\Bigl(\frac{a}{t}\Bigr) = a \lim_{u \rightarrow \infty} \frac{f(u)}{u}, \quad \forall \, a > 0.
\end{align*}
\label{definition:f-divergence}
\end{definition}

\begin{definition}
An $f$-divergence is said to be {\em symmetric} if $D_f(P||Q) = D_f(Q||P)$ for every $P$ and $Q$.
\end{definition}

Symmetric $f$-divergences include (among others) the squared Hellinger distance where $$f(t) = (\sqrt{t}-1)^2, \quad
D_f(P||Q) = \sum_x \left(\sqrt{P(x)} - \sqrt{Q(x)}\right)^2,$$ and the total variation distance in
\eqref{eq: the L1 distance is twice the total variation distance} where $f(t) = \frac{1}{2} \, |t-1|.$

An $f$-divergence is symmetric if and only if the function $f$ satisfies the equality (see \cite[p.~765]{Gilardoni06})
\begin{equation}
f(u) = u \, f\left(\frac{1}{u}\right) + a (u-1), \quad \forall \, u \in (0, \infty)
\label{eq: characterization of f of a symmetric f-divergence}
\end{equation}
for some constant $a$. If $f$ is differentiable at $u=1$ then a differentiation of both sides of equality
\eqref{eq: characterization of f of a symmetric f-divergence} at $u=1$ gives that $a = 2 f'(1)$.

Note that the relative entropy (a.k.a. the Kullback-Leibler divergence)
$D(P||Q) \triangleq \sum_{x} P(x) \log\left(\frac{P(x)}{Q(x)}\right)$
is an $f$-divergence with $f(t) = t \log(t), \; t>0$; its dual, $D(Q||P)$, is an f-divergence with
$f(t) = -\log(t), \; t>0$; clearly, it is an asymmetric $f$-divergence since $D(P||Q) \neq D(Q||P)$ .

\vspace*{0.1cm}
The following result, which was derived by Gilardoni (see \cite{Gilardoni06, Gilardoni10}),
refers to the infimum of a symmetric $f$-divergence for a fixed value of the total variation distance:
\begin{theorem}
Let $f \colon (0, \infty) \rightarrow \reals$ be a convex function with $f(1)=0$,
and assume that $f$ is twice differentiable. Let
\begin{equation*}
L_{D_f}(\varepsilon) \triangleq \inf_{P, Q \colon \, d_{\text{TV}}(P,Q) = \varepsilon} D_f(P||Q),
\quad \forall \, \varepsilon \in [0,1]
\end{equation*}
be the infimum of the $f$-divergence for a given total variation distance. If $D_f$
is a symmetric $f$-divergence, and $f$ is differentiable at $u=1$, then
\begin{equation*}
L_{D_f}(\varepsilon) = (1-\varepsilon) \, f\left(\frac{1+\varepsilon}{1-\varepsilon}\right)
- 2 f'(1) \, \varepsilon, \quad \forall \, \varepsilon \in [0,1].
\end{equation*}
\label{theorem: lower bound on symmetric f-divergence in terms of the total variation distance}
\end{theorem}

\section{Tight Bounds on Symmetric Divergence Measures}
\label{section: Derivation of Tight Bounds on Symmetric Distance Measures}
The following section introduces tight bounds for several symmetric divergence
measures (where part of them are not $f$-divergences) for a fixed value of the
total variation distance.

\subsection{Tight Bounds on the Bhattacharyya Coefficient}
\label{subsection: Tight Bounds on the Bhattacharyya Coefficient}
\begin{definition}
Let  $P$ and $Q$ be two probability distributions that are defined on the same
set. The {\em Bhattacharyya coefficient} between $P$ and $Q$ is given by
$Z(P, Q) \triangleq \sum_{x} \sqrt{P(x) \, Q(x)}$.
\label{definition: probability metrics}
\end{definition}

\begin{proposition}
Let $P$ and $Q$ be two probability distributions. Then, for a fixed value
$\varepsilon \in [0,1]$ of the total variation distance (i.e., if
$d_{\text{TV}}(P,Q)=\varepsilon$), the respective Bhattacharyya
coefficient satisfies the inequality
$1-\varepsilon \leq Z(P, Q) \leq \sqrt{1 - \varepsilon^2}$.
Both upper and lower bounds are tight: the upper bound is attained by
the pair of 2-element probability distributions
$P = \left( \frac{1-\varepsilon}{2}, \, \frac{1+\varepsilon}{2} \right)$, and
$Q = \left( \frac{1+\varepsilon}{2}, \, \frac{1-\varepsilon}{2} \right),$ and the
lower bound is attained by the pair of 3-element probability distributions
$P=(\varepsilon, 1-\varepsilon, 0)$, and $Q = (0, 1-\varepsilon, \varepsilon).$
\label{proposition: tight bounds on the Bhattacharyya Coefficient for a given total variation distance}
\end{proposition}

\begin{remark}
Although derived independently in this work,
Proposition~\ref{proposition: tight bounds on the Bhattacharyya Coefficient for a given total variation distance}
is a known result in quantum information theory (on the
relation between the trace distance and fidelity \cite{wikipedia}).
\end{remark}

\subsection{A Tight Bound on the Chernoff Information}
\label{subsection: A Tight Bound on the Chernoff Information}
\begin{definition}
The {\em Chernoff information} between two probability distributions $P$ and $Q$,
defined on the same set, is
\begin{align*}
& C(P,Q) \triangleq -\min_{\lambda \in [0,1]} \; \log \left( \sum_{x}
P(x)^{\lambda} \, Q(x)^{1-\lambda} \right)
\end{align*}
where throughout this paper, the logarithms are on base~$e$.
\label{definition: Chernoff information}
\end{definition}

\begin{proposition}
Let
\begin{align}
C(\varepsilon) \triangleq \min_{P, Q \colon \, d_{\text{TV}}(P,Q) = \varepsilon} C(P, Q),
\quad \forall \, \varepsilon \in [0,1]
\label{eq: minimal Chernoff information for given TV distance}
\end{align}
be the minimum of the Chernoff information for a fixed value $\varepsilon \in [0,1]$ of the
total variation distance.
This minimum indeed exists, and it is equal to
\begin{equation*}
C(\varepsilon) = \left\{\begin{array}{ll}
                 -\frac{1}{2} \, \log(1-\varepsilon^2)   & \mbox{if $\varepsilon \in [0,1)$} \\[0.1cm]
                 +\infty                                 & \mbox{if $\varepsilon = 1$.}
\end{array}
\right.
\end{equation*}
For $\varepsilon \in [0,1)$, it is achieved by
the pair of 2-element probability distributions
$P = \left( \frac{1-\varepsilon}{2}, \, \frac{1+\varepsilon}{2} \right)$, and
$Q = \left( \frac{1+\varepsilon}{2}, \, \frac{1-\varepsilon}{2} \right)$.
\label{proposition: Chernoff information}
\end{proposition}

{\em Outline of the proof}: Definition~\ref{definition: Chernoff information} with a
possibly sub-optimal value of $\lambda=\frac{1}{2}$, and
Proposition~\ref{proposition: tight bounds on the Bhattacharyya Coefficient for a given total variation distance}
yield that
\begin{align}
C(P,Q) & \geq -\log \left(\sum_x \sqrt{P(x) \, Q(x)}\right) \nonumber \\
& = -\log Z(P,Q) \nonumber \\
& \geq -\frac{1}{2} \, \Bigl( 1- \bigl(d_{\text{TV}}(P,Q) \bigr)^2 \Bigr).
\label{eq: lower bound on the Chernoff information}
\end{align}
Consequently, from \eqref{eq: minimal Chernoff information for given TV distance},
$C(\varepsilon) \geq -\frac{1}{2} \log(1-\varepsilon^2)$ for $\varepsilon \in [0,1)$.
It can be verified that the lower bound on $C(P,Q)$ is achieved for
$P = \left( \frac{1-\varepsilon}{2}, \, \frac{1+\varepsilon}{2} \right)$, and
$Q = \left( \frac{1+\varepsilon}{2}, \, \frac{1-\varepsilon}{2} \right)$.

\vspace*{0.1cm}
\begin{remark} A geometric interpretation of the minimum of the Chernoff information subject to
a minimal total variation distance has been recently provided in \cite[Section~3]{Sason_arXiv15}.
\end{remark}

\vspace*{0.1cm}
\begin{remark}[An Application]
From \eqref{eq: lower bound on the Chernoff information},
a lower bound on the total variation distance implies a lower bound on the Chernoff
information; consequently, it provides an upper bound on the best achievable Bayesian probability
of error for binary hypothesis testing.
This approach has been recently used in \cite{YardiKV14} to obtain a lower bound on the Chernoff
information for studying a communication problem that is related to channel-code detection via
the likelihood ratio test.
\end{remark}

\subsection{A Tight Bound on the Capacitory Discrimination}
\label{subsection: A Tight Bound on the Capacitory Discrimination}
\label{subsection: Bounds on the Capacitory Discrimination}
The capacitory discrimination (a.k.a. the Jensen-Shannon divergence) is defined as follows:
\begin{definition}
Let $P$ and $Q$ be two probability distributions.
The capacitory discrimination between $P$ and $Q$ is given by
\begin{equation*}
\begin{split}
\overline{C}(P,Q) &\ \triangleq D\left(P \, || \,
\frac{P+Q}{2}\right)+D\left(Q \, || \, \frac{P+Q}{2}\right) \\
&\ = 2 \left[H\left(\frac{P+Q}{2}\right) - \frac{H(P)+H(Q)}{2} \right].
\end{split}
\end{equation*}
\label{definition: capacitory discrimination}
\end{definition}
This divergence measure was studied, e.g., in \cite{Lin91}
and \cite{Topsoe_IT00}.

\begin{proposition}
For every $\varepsilon \in [0,1)$,
\begin{equation}
\min_{P, Q \colon \, d_{\text{TV}}(P,Q) = \varepsilon}
\overline{C}(P,Q) = 2 \, d\left(\frac{1-\varepsilon}{2} \,
\big|\big| \, \frac{1}{2} \right)
\label{eq: a tight lower bound on the capacitory discrimination}
\end{equation}
and it is achieved by the 2-element probability distributions
$P = \left( \frac{1-\varepsilon}{2}, \, \frac{1+\varepsilon}{2} \right)$, and
$Q = \left( \frac{1+\varepsilon}{2}, \, \frac{1-\varepsilon}{2} \right)$. In
\eqref{eq: a tight lower bound on the capacitory discrimination},
$$d(p||q) \triangleq p \log\left(\frac{p}{q}\right)+
(1-p)\log\left(\frac{1-p}{1-q}\right), \quad p, q \in [0,1],$$
with the convention that $0 \log 0 =0$.
\label{proposition: minimum of the capacitory discrimination for a fixed total variation distance}
\end{proposition}

{\em Outline of the proof}: In \cite[p.~119]{GSS_IT14}, $\overline{C}(P,Q)=D_f(P\|Q)$
with $ f(t) = t \log t - (t+1) \log(1+t) + 2 \log 2$ for $t > 0$. This is a symmetric
$f$-divergence where $f$ is convex with $f(1)=0$, $f'(1)=-\log 2$.
Eq.~\eqref{eq: a tight lower bound on the capacitory discrimination}
follows from Theorem~\ref{theorem: lower bound on symmetric f-divergence in terms of the total variation distance}.

\subsection{A Tight Bound on Jeffreys' divergence}
\label{subsection: Bounds on Jeffreys' divergence}
\begin{definition}
Let $P$ and $Q$ be two probability distributions. Jeffreys' divergence \cite{Jeffreys46}
is a symmetrized version of the relative entropy, which is defined as
\begin{equation}
J(P,Q) \triangleq \frac{D(P||Q) + D(Q||P)}{2}.
\label{eq: Jeffreys' divergence}
\end{equation}
\end{definition}

\begin{proposition}
For every $\varepsilon \in [0,1)$,
\begin{align}
\min_{P,Q \colon d_{\text{TV}}(P,Q) = \varepsilon} J(P,Q)
= \varepsilon \, \log\left(\frac{1+\varepsilon}{1-\varepsilon}\right).
\label{eq: a tight lower bound on Jeffreys' divergence in terms of the total variation distance}
\end{align}
The minimum in \eqref{eq: a tight lower bound on Jeffreys' divergence in terms of the total variation distance}
is achieved by the pair of 2-element distributions
$P = \left( \frac{1-\varepsilon}{2}, \, \frac{1+\varepsilon}{2} \right)$ and
$Q = \left( \frac{1+\varepsilon}{2}, \, \frac{1-\varepsilon}{2} \right)$.
\label{proposition: tight lower bound on Jeffreys' divergence in terms of the total variation distance}
\end{proposition}

{\em Outline of the proof}: Jeffreys' divergence can be expressed
as a symmetric $f$-divergence where $f(t) = \frac{1}{2} \, (t-1) \log t$ for $t>0$.
Note that $f$ is convex, and $f(1)= f'(1)=0$.
Eq.~\eqref{eq: a tight lower bound on Jeffreys' divergence in terms of the total variation distance}
follows from Theorem~\ref{theorem: lower bound on symmetric f-divergence in terms of the total variation distance}.

\section{A Bound for Lossless Source Coding}
\label{section: bound for lossless source coding}

We illustrate in the following a use of
Proposition~\ref{proposition: tight lower bound on Jeffreys' divergence in terms of the total variation distance}
for lossless source coding. This tightens,
and also refines under a certain condition,
a bound by Csisz\'{a}r \cite{Csiszar67b}.

Consider a memoryless and stationary source with alphabet
$\mathcal{U}$ that emits symbols according to a probability
distribution $P$, and assume a uniquely decodable (UD) code
with an alphabet of size $d$. It is well known that such a
UD code achieves the entropy of the source if and only if
the length $l(u)$ of the codeword that is assigned to each
symbol $u \in \mathcal{U}$ satisfies the equality
$l(u) = -\log_d P(u)$ for every $u \in \mathcal{U}$.
This corresponds to a dyadic source where, for every
$u \in \mathcal{U}$, we have $P(u) = d^{-n_u}$
with a natural number $n_u$; in this case, $l(u) = n_u$
for every symbol $u \in \mathcal{U}$. Let
$\overline{L} \triangleq \expectation[L]$ designate the
average length of the codewords, and
$H_d(U) \triangleq -\sum_{u \in \mathcal{U}} P(u) \, \log_d P(u)$
be the entropy of the source (to the base $d$). Furthermore, let
$c_{d,l} \triangleq \sum_{u \in \mathcal{U}} d^{-l(u)}.$
According to the Kraft-McMillian inequality, the inequality
$c_{d,l} \leq 1$ holds in general for UD codes,
and the equality $c_{d,l}=1$ holds if
the code achieves the entropy of the source (i.e.,
$\overline{L} = H_d(U)$).

Define the probability distribution
$Q_{d,l}(u) \triangleq \left(\frac{1}{c_{d,l}} \right) \, d^{-l(u)}$
for every $u \in \mathcal{U}$,
and let $\Delta_d \triangleq \overline{L} - H_d(U)$ designate the
redundancy of the code. Note that for a UD code that achieves
the entropy of the source, its probability distribution $P$ is
equal to $Q_{d,l}$ (since $c_{d,l}=1$, and $P(u) = d^{-l(u)}$
for every $u \in \mathcal{U}$).

In \cite{Csiszar67b}, a generalization for UD source codes has been
studied by a derivation of an upper bound on the $L_1$ norm between
the two probability distributions $P$ and $Q_{d,l}$ as a function
of the redundancy $\Delta_d$ of the code. To this end, straightforward
calculation shows that the relative entropy from $P$ to $Q_{d,l}$ is given by
\begin{equation}
D(P || Q_{d,l}) = \Delta_d \, \log d + \log \bigl(c_{d,l}\bigr).
\label{eq:relative entropy from P to Q}
\end{equation}
The interest in \cite{Csiszar67b} is in getting an upper bound that
only depends on the (average) redundancy $\Delta_d$ of the code, but
is independent of the specific distribution of the length of each codeword.
Hence, since the Kraft-McMillian inequality states that $c_{d,l} \leq 1$
for general UD codes, it is concluded in \cite{Csiszar67b} that
\begin{equation}
D(P || Q_{d,l}) \leq \Delta_d \, \log d.
\label{eq:upper bound on the relative entropy}
\end{equation}
Consequently, it follows from Pinsker's inequality that
\begin{equation}
\sum_{u \in \mathcal{U}} \bigl| P(u) - Q_{d,l}(u) \bigr| \leq \min
\bigl\{ \sqrt{2 \Delta_d \log d}, \, 2 \bigr\}
\label{eq:Csiszar's bound for lossless source coding}
\end{equation}
where it is also taken into account that, from the triangle inequality,
the sum on the left-hand side of
\eqref{eq:Csiszar's bound for lossless source coding} cannot exceed~2.
This inequality is indeed consistent with the fact that the probability
distributions $P$ and $Q_{d,l}$ coincide when $\Delta_d = 0$ (i.e., for
a UD code which achieves the entropy of the source).

At this point we deviate from the analysis in \cite{Csiszar67b}.
One possible improvement of the bound in \eqref{eq:Csiszar's bound for lossless source coding}
follows by replacing Pinsker's inequality with the result in
\cite{FedotovHT_IT03}, i.e., by taking into account the exact
parametrization of the infimum of the relative entropy for a
given total variation distance. This gives the following tightened bound:
\begin{equation}
\sum_{u \in \mathcal{U}} \bigl| P(u) - Q_{d,l}(u) \bigr| \leq 2 \;
L^{-1}(\Delta_d \log d)
\label{eq:first tightening of Csiszar's bound for lossless source coding}
\end{equation}
where $L^{-1}$ is the inverse function of $L$, given as follows \cite{ReidW11}:
\begin{align}
L(\varepsilon) & \triangleq \inf_{P, Q \colon \, d_{\text{TV}}(P,Q)
= \varepsilon} D(P||Q) \nonumber \\
& = \min_{\beta \in [\varepsilon-1, \, 1-\varepsilon]} \left\{
\left(\frac{\varepsilon+1-\beta}{2}\right) \, \log
\left(\frac{\beta-1-\varepsilon}{\beta-1+\varepsilon}\right) \right.
\nonumber \\[0.1cm]
& \hspace*{2cm} \left. + \left(\frac{\beta+1-\varepsilon}{2}\right)
\, \log\left(\frac{\beta+1-\varepsilon}{\beta+1+\varepsilon}\right) \right\}.
\label{eq:Reid and Willimason's bound for the relative entropy}
\end{align}
It can be verified that the numerical minimization w.r.t. $\beta$ in
\eqref{eq:Reid and Willimason's bound for the relative entropy} can be restricted
to the interval $[\varepsilon-1, \, 0]$ (it is calculated numerically).

In the following, the utility of
Proposition~\ref{proposition: tight lower bound on Jeffreys' divergence in terms of the total variation distance}
is shown by refining the bound in
\eqref{eq:first tightening of Csiszar's bound for lossless source coding}. Let
$\delta(u) \triangleq l(u) + \log_d P(u)$ for every $u \in \mathcal{U}.$
Calculation of the dual divergence gives
\begin{align}
D(Q_{d,l} || P)
= -\log \bigl(c_{d,l}\bigr) - \left(\frac{\log d}{c_{d,l}} \right)
\expectation \bigl[\delta(U) \, d^{-\delta(U)}\bigr]
\label{eq:relative entropy from Q to P}
\end{align}
and the combination of \eqref{eq: Jeffreys' divergence}, \eqref{eq:relative entropy from P to Q}
and \eqref{eq:relative entropy from Q to P} yields that
{\small \begin{equation}
J(P, Q_{d,l}) = \frac{1}{2} \left[\Delta_d \log d - \left(\frac{\log d}{c_{d,l}}\right)
\expectation \bigl[\delta(U) \, d^{-\delta(U)}\bigr] \right].
\label{eq:Jeffreys' divergence between P and Q}
\end{equation}}
For the simplicity of the continuation of the analysis, we restrict our attention to UD codes
that satisfy the condition
\begin{equation}
l(u) \geq \left\lceil \log_d \frac{1}{P(u)} \right\rceil, \quad \forall \, u \in \mathcal{U}.
\label{eq:condition for further sarpening the bound for lossless source coding}
\end{equation}
In general, it excludes Huffman codes; nevertheless, it is
satisfied by some other important UD codes such as the Shannon
code, Shannon-Fano-Elias code, and arithmetic coding. Since
\eqref{eq:condition for further sarpening the bound for lossless source coding}
is equivalent to the condition that $\delta$ is non-negative on
$\mathcal{U}$, it follows from
\eqref{eq:Jeffreys' divergence between P and Q} that
\begin{equation}
J(P, Q_{d,l}) \leq \frac{\Delta_d \log d}{2}
\label{eq:upper bound on Jeffreys' divergence}
\end{equation}
so, the upper bound on Jeffreys' divergence in
\eqref{eq:upper bound on Jeffreys' divergence} is
twice smaller than the upper bound on the relative
entropy in \eqref{eq:upper bound on the relative entropy}.
It is partially because the term $\log c_{d,l}$ is canceled
out along the derivation of the bound in
\eqref{eq:upper bound on Jeffreys' divergence},
in contrast to the derivation of the bound in
\eqref{eq:upper bound on the relative entropy}
where this term was removed from the bound in order
to avoid its dependence on the length of the codeword
for each individual symbol.

Following Proposition~\ref{proposition: tight lower bound on Jeffreys' divergence in terms of the total variation distance},
for an arbitrary $x \geq 0$, let $\varepsilon \triangleq \varepsilon(x)$
be the solution in the interval $[0, 1)$ of the equation
\begin{equation}
\varepsilon \, \log\left(\frac{1+\varepsilon}{1-\varepsilon}\right) = x.
\label{eq:equation for J divergence}
\end{equation}
The combination of \eqref{eq: a tight lower bound on Jeffreys' divergence in terms of the total variation distance}
and \eqref{eq:upper bound on Jeffreys' divergence} implies that
\begin{equation}
\sum_{u \in \mathcal{U}} \bigl| P(u) - Q_{d,l}(u) \bigr|
\leq 2 \; \varepsilon \left( \frac{\Delta_d \log d}{2} \right).
\label{eq:second tightening of Csiszar's bound for lossless source coding}
\end{equation}

In the following, the bounds in
\eqref{eq:first tightening of Csiszar's bound for lossless source coding}
and \eqref{eq:second tightening of Csiszar's bound for lossless source coding}
are compared analytically for the case where the average redundancy is small
(i.e., $\Delta_d \approx 0$). Under this approximation, the bound in
\eqref{eq:Csiszar's bound for lossless source coding} (i.e., the original bound
from \cite{Csiszar67b}) coincides with its tightened version in
\eqref{eq:first tightening of Csiszar's bound for lossless source coding}.
On the other hand, since for $\varepsilon \approx 0$, the left-hand side of
\eqref{eq:equation for J divergence} is approximately $2 \varepsilon^2$,
it follows from \eqref{eq:equation for J divergence} that, for $x \approx 0$,
we have $\varepsilon(x) \approx \sqrt{\frac{x}{2}}.$
It follows that, if $\Delta_d \approx 0$, inequality
\eqref{eq:second tightening of Csiszar's bound for lossless source coding} gets
approximately the form
$$\sum_{u \in \mathcal{U}} \bigl| P(u) - Q_{d,l}(u) \bigr| \leq \sqrt{\Delta_d \log d}.$$
Hence, even for a small redundancy, the bound in
\eqref{eq:second tightening of Csiszar's bound for lossless source coding}
improves \eqref{eq:Csiszar's bound for lossless source coding} by a factor of $\sqrt{2}$.

A numerical comparison of the bounds in
\eqref{eq:Csiszar's bound for lossless source coding},
\eqref{eq:first tightening of Csiszar's bound for lossless source coding} and
\eqref{eq:second tightening of Csiszar's bound for lossless source coding} is
provided in the journal paper, see \cite[Figure~2]{Sason_IT15}.

\begin{remark}
Another application of Jeffreys' divergence has been recently
studied in \cite[Section~5]{Arjmandi} where the mutual information
$I(X;Y) = D(P_{X,Y} \| P_X P_Y)$ has been upper bounded by the symmetrized divergence
\begin{align*}
D_{\text{sym}}(P_{X,Y} \| P_X P_Y) &= D(P_{X,Y} \| P_X P_Y) + D(P_X P_Y \| P_{X,Y}) \\
&= 2 J(P_{X,Y}, P_X P_Y).
\end{align*}
Consequently, the channel capacity satisfies the upper bound
$C = \max_{P_X} I(X;Y) \leq 2 \max_{P_X} J(P_{X,Y}, P_X P_Y).$
This provides a good bound on the channel
capacity in the low SNR regime (see \cite[Section~5]{Arjmandi}).
It has been applied in \cite[Section~6]{Arjmandi} to obtain
a bound on the capacity of a linear-time invariant
Poisson channel; this bound is improved by increasing
the parameter of the background noise $(\lambda_0)$ \cite{Arjmandi}.
\end{remark}

\section{A New Inequality Relating $f$-Divergences}
\label{section: a new inequality relating f-divergences}
We introduce in the following an inequality which relates $f$-divergences,
and its use is exemplified. This inequality is proved here
since it is not included in the journal paper \cite{Sason_IT15}.

Recall the following definition of the $\chi^2$-divergence.
\begin{definition}
The {\em chi-squared divergence} between two probability distributions $P$ and
$Q$ on a set $\mathcal{A}$ is given by
\begin{equation}
\chi^2(P,Q) \triangleq \sum_{x \in \mathcal{A}} \frac{\bigl(P(x)-Q(x)\bigr)^2}{Q(x)}
= \sum_{x \in \mathcal{A}} \frac{P(x)^2}{Q(x)} - 1 \, .
\label{eq:chi-squared divergence}
\end{equation}
\end{definition}
The chi-squared divergence is an asymmetric $f$-divergence where $f(t) = (t-1)^2$ for $t \geq 0$.

\begin{proposition}
Let $f \colon (0, \infty) \rightarrow \reals$ be a convex function with $f(1)=0$ and further
assume that the function $g \colon (0, \infty) \rightarrow \reals$, defined by
$g(t)=-t f(t)$ for every $t>0$, is also convex. Let $P$ and $Q$ be two probability
distributions on a finite set $\mathcal{A}$, and assume that $P, Q$ are positive on this set.
Then, the following inequality holds:
\begin{align}
& \min_{x \in \mathcal{A}} \frac{P(x)}{Q(x)} \cdot D_f(P||Q) \nonumber \\
& \leq -D_g(P||Q) - f\bigl(1 + \chi^2(P,Q)\bigr) \nonumber \\
& \leq \max_{x \in \mathcal{A}} \frac{P(x)}{Q(x)} \cdot D_f(P||Q).
\label{eq:new inequality relating f-divergences}
\end{align}
\label{proposition:new inequality relating f-divergences}
\end{proposition}

\vspace*{-0.5cm}
\begin{proof}
Let $\mathcal{A} = \bigl\{x_1, \ldots, x_n \bigr\}$, and
$\underline{u} = (u_1, \ldots, u_n) \in \reals_+^n$ be an arbitrary $n$-tuple
with positive entries. Define
\begin{align}
\begin{split}
& J_n(f, \underline{u}, P) \triangleq \sum_{i=1}^n P(x_i) \, f(u_i)
- f\left(\sum_{i=1}^n P(x_i) u_i \right), \\[0.1cm]
& J_n(Q, \underline{u}, P) \triangleq \sum_{i=1}^n Q(x_i) \, f(u_i)
- f\left(\sum_{i=1}^n Q(x_i) u_i \right).
\label{eq:Jensen functional}
\end{split}
\end{align}
The following refinement of Jensen's inequality appears in
\cite[Theorem~1]{Dragomir06} for a convex function $f \colon (0, \infty) \rightarrow \reals$,
and it has been extended in \cite[Theorem~1]{BaricM09} to hold for a convex $f$ over an
arbitrary interval $[a,b]$:
\begin{align}
& \min_{i \in \{1, \ldots, n\}} \left\{\frac{P(x_i)}{Q(x_i)} \right\} \, J_n(f, \underline{u}, Q)
\leq J_n(f, \underline{u}, P) \nonumber \\
& \leq \max_{i \in \{1, \ldots, n\}} \left\{ \frac{P(x_i)}{Q(x_i)} \right\} \, J_n(f, \underline{u}, Q).
\label{eq:Dragomir's inequality ('06)}
\end{align}
The refined version of Jensen's inequality in \eqref{eq:Dragomir's inequality ('06)} is applied
in the following to prove \eqref{eq:new inequality relating f-divergences}. Let
$u_i \triangleq \frac{P(x_i)}{Q(x_i)}$ for $i \in \{1, \ldots, n\}$.
Calculation of \eqref{eq:Jensen functional} gives that
{\small \begin{align}
J_n(f, \underline{u}, Q)
& = \sum_{i=1}^n Q(x_i) \, f\left(\frac{P(x_i)}{Q(x_i)}\right) - f\left(\sum_{i=1}^n Q(x_i)
\cdot \frac{P(x_i)}{Q(x_i)} \right) \nonumber \\
& = \sum_{x \in \mathcal{A}} Q(x) \, f\left(\frac{P(x)}{Q(x)}\right) - f(1)
= D_f(P||Q), \label{eq:J1} \\[0.1cm]
J_n(f, \underline{u}, P)
& = \sum_{i=1}^n P(x_i) \, f\left(\frac{P(x_i)}{Q(x_i)}\right)
- f\left(\sum_{i=1}^n \frac{P(x_i)^2}{Q(x_i)} \right) \nonumber \\
& \stackrel{(\text{a})}{=} - \sum_{i=1}^n Q(x_i) \, g\left(\frac{P(x_i)}{Q(x_i)}\right)
- f\left(\sum_{i=1}^n \frac{P(x_i)^2}{Q(x_i)} \right) \nonumber \\
& \stackrel{(\text{b})}{=} - D_g(P||Q) - f\bigl(1+\chi^2(P,Q)\bigr)
\label{eq:J2}
\end{align}}
where equality~(a) holds by the definition of $g$, and equality~(b) follows from
equalities \eqref{eq:f-divergence} and \eqref{eq:chi-squared divergence}. The
substitution of \eqref{eq:J1} and \eqref{eq:J2} in \eqref{eq:Dragomir's inequality ('06)}
completes the proof.
\end{proof}

As a consequence of Proposition~\ref{proposition:new inequality relating f-divergences},
we introduce the following inequality which relates between the relative entropy,
its dual and the chi-squared divergence.

\begin{corollary}
Let $P$ and $Q$ be two probability distributions on a finite set $\mathcal{A}$,
and assume that $P, Q$ are positive on $\mathcal{A}$. Then, the following inequality holds:
\begin{align}
& \min_{x \in \mathcal{A}} \frac{P(x)}{Q(x)} \cdot D(Q||P) \nonumber \\
& \leq \log \bigl( 1 + \chi^2(P,Q) \bigr) - D(P||Q) \nonumber \\
& \leq \max_{x \in \mathcal{A}} \frac{P(x)}{Q(x)} \cdot D(Q||P).
\label{eq: relation between relative entropy, dual of relative entropy and chi-squared divergence}
\end{align}
\label{corollary: relation between relative entropy, dual of relative entropy and chi-squared divergence}
\end{corollary}

\vspace*{-0.5cm}
\begin{proof}
Let $f(t) = -\log(t)$ for $t>0$. The function $f \colon (0, \infty) \rightarrow \reals$
is convex with $f(1)=0$, and $g(t) = -t f(t) = t \log(t)$ for $t>0$ defines a
convex function with $g(1)=0$. Inequality
\eqref{eq: relation between relative entropy, dual of relative entropy and chi-squared divergence}
follows by substituting $f, g$ in \eqref{eq:new inequality relating f-divergences}
where $D_f(P||Q) = D(Q||P)$ and $D_g(P||Q) = D(P||Q)$.
\end{proof}

\vspace*{0.1cm}
\begin{remark}
Inequality~\eqref{eq: relation between relative entropy, dual of relative entropy and chi-squared divergence}
strengthens the inequality
\begin{equation}
\chi^2(P,Q) \geq e^{D(P||Q)}-1
\label{eq:lower bound of the chi-squared divergence in terms of the divergence}
\end{equation}
which is derived by using Jensen's inequality as follows \cite{DragomirG_01}:
\begin{align*}
\chi^2(P,Q) 
&= \sum_{x \in \mathcal{A}} \Bigl\{ P(x) e^{\log \left(\frac{P(x)}{Q(x)}\right)} \Bigr\} - 1 \\
&\geq e^{\sum_{x \in \mathcal{A}} P(x) \, \log \left(\frac{P(x)}{Q(x)}\right)} - 1 \\
&= e^{D(P||Q)}-1.
\end{align*}
\end{remark}

The following inequality is another consequence of
Proposition~\ref{proposition:new inequality relating f-divergences},
relating the chi-squared divergence and its dual:
\vspace*{0.1cm}
\begin{corollary}
Under the same conditions of
Corollary~\ref{corollary: relation between relative entropy, dual of relative entropy and chi-squared divergence},
the following inequality holds:
{\small \begin{align*}
\min_{x \in \mathcal{A}} \frac{P(x)}{Q(x)} \cdot \chi^2(Q,P)
\leq \frac{\chi^2(P,Q)}{1+\chi^2(P,Q)}
\leq \max_{x \in \mathcal{A}} \frac{P(x)}{Q(x)} \cdot \chi^2(Q,P).
\end{align*}}
\label{corollary: chi-squared divergence and its dual divergence}
\end{corollary}
\vspace*{-0.5cm}
\begin{proof}
This follows from Proposition~\ref{proposition:new inequality relating f-divergences}
where $f(t) = \frac{1}{t}-1$, and $g(t) = -t f(t) = t-1$ for $t>0$.
Consequently, we have $D_g(P||Q) = 0$,
$D_f(P||Q) = \chi^2(Q,P)$.
\end{proof}

\vspace*{-0.2cm}
\subsection*{Acknowledgment}
The author thanks the anonymous reviewers of the journal paper in the {\em IEEE Trans.
on Information Theory} \cite{Sason_IT15}, and of this conference paper at the proceedings
of the 2015 IEEE Information Theory Workshop in Israel for their helpful comments.
This research work has been supported by the Israeli Science Foundation (ISF), grant number 12/12.

\end{document}